\documentclass[conference]{IEEEtran}
\IEEEoverridecommandlockouts
\usepackage{cite}
\usepackage{amsmath,amssymb,amsfonts}
\usepackage{textcomp}
\usepackage{xcolor}

\usepackage[printonlyused]{acronym}
\usepackage{booktabs} 
\usepackage{multirow}
\usepackage{graphicx}
\usepackage{caption}
\usepackage{subcaption}
\usepackage[ruled,vlined]{algorithm2e}
\usepackage{tipa}
\usepackage{siunitx}
\usepackage{color,soul}

\acrodef{GWAS}[GWAS]{Genome-Wide Association Studies}
\acrodef{SNP}[SNP]{Single-Nucleotide Polymorphism}
\acrodef{iGPU}[iGPU]{integrated Graphics Processing Unit}
\acrodef{GPU}[GPU]{Graphics Processing Unit}
\acrodef{FPGA}[FPGA]{Field-Programmable Gate Array}
\acrodef{CPU}[CPU]{Central Processing Unit}
\acrodef{AVX}[AVX]{Advanced Vector Extension}
\acrodef{AVX256}[AVX256]{256-bits Advanced Vector Extension}
\acrodef{AVX512}[AVX512]{512-bits Advanced Vector Extension}
\acrodef{SSE}[SSE]{Streaming SIMD Extension}
\acrodef{LLC}[LLC]{Last Level Cache}
\acrodef{DRAM}[DRAM]{Dynamic Random Access Memory}
\acrodef{EU}[EU]{Execution Unit}
\acrodef{ALU}[ALU]{Arithmetic Logic Unit}
\acrodef{FP}[FP]{Floating-Point}
\acrodef{OpenMP}[OpenMP]{Open Multi-Processing}
\acrodef{OpenCL}[OpenCL]{Open Compute Language}
\acrodef{PAPI}[PAPI]{Performance API}
\acrodef{RAPL}[RAPL]{Running Average Power Limit}
\acrodef{SIMD}[SIMD]{Single Instruction Multiple Data}
\acrodef{CUDA}[CUDA]{Compute Unified Device Architecture}
\acrodef{DPC++}[DPC++]{Data-Parallel C++}
\acrodef{MPI}[MPI]{Message Passing Interface}
\acrodef{OOO}[OoO]{Out-of-Order}
\acrodef{SLM}[SLM]{Shared Local Memory}

\acrodef{API}[API]{Application Programming Interface}
\acrodef{DVFS}[DVFS]{Dynamic Voltage and Frequency Scaling}
\acrodef{EDP}[EDP]{Energy Delay Product}
\acrodef{ISA}[ISA]{Instruction Set Architecture}
\acrodef{OS}[OS]{Operating System}
\acrodef{PARSEC}[PARSEC]{Princeton Application Repository for Shared-Memory Computers}
\acrodef{PMU}[PMU]{Performance Monitoring Unit}
\acrodef{RISC}[RISC]{Reduced Instruction Set Computer}
\acrodef{SOC}[SoC]{System on a Chip}
\acrodef{TDP}[TDP]{Thermal Design Power}
\acrodef{CARM}[CARM]{Cache-Aware Roofline Model}
\acrodef{SM}[SM]{Stream Multiprocessor}
\acrodef{AI}[AI]{Arithmetic Intensity}

\def\BibTeX{{\rm B\kern-.05em{\sc i\kern-.025em b}\kern-.08em
    T\kern-.1667em\lower.7ex\hbox{E}\kern-.125emX}}
    
\begin{document}

\title{Unlocking Personalized Healthcare on Modern CPUs/GPUs: Three-way Gene Interaction Study\\
}

\author{
\IEEEauthorblockN{Diogo Marques\IEEEauthorrefmark{1},
Rafael Campos\IEEEauthorrefmark{1},
Sergio~Santander-Jiménez\IEEEauthorrefmark{2},
Zakhar Matveev\IEEEauthorrefmark{4},
Leonel~Sousa\IEEEauthorrefmark{1} and Aleksandar~Ilic\IEEEauthorrefmark{1}}
\IEEEauthorblockA{\IEEEauthorrefmark{1}INESC-ID\\
Instituto Superior Técnico, Universidade de Lisboa, Portugal\\
Email: \{diogo.marques, leonel.sousa, aleksandar.ilic\}@inesc-id.pt and rafaeltcampos@tecnico.ulisboa.pt}
\IEEEauthorblockA{\IEEEauthorrefmark{2}Polytechnic School, University of Extremadura, Spain\\
Email: sesaji@unex.es}
\IEEEauthorblockA{\IEEEauthorrefmark{4}Intel Corporation, Russia\\
Email: zakhar.a.matveev@intel.com}}

\maketitle

\thispagestyle{plain}
\pagestyle{plain}

\begin{abstract}
Developments in Genome-Wide Association Studies have led to the increasing notion that future healthcare techniques will be personalized to the patient, by relying on genetic tests to determine the risk of developing a disease. To this end, the detection of gene interactions that cause complex diseases constitutes an important application. Similarly to many applications in this field, extensive data sets containing genetic information for a series of patients are used (such as Single-Nucleotide Polymorphisms), leading to high computational complexity and memory utilization, thus constituting a major challenge when targeting high-performance execution in modern computing systems. To close this gap, this work proposes several novel approaches for the detection of three-way gene interactions in modern CPUs and GPUs, making use of different optimizations to fully exploit the target architectures. Crucial insights from the Cache-Aware Roofline Model are used to ensure the suitability of the applications to the computing devices. An extensive study of the architectural features of 13 CPU and GPU devices from all main vendors is also presented, allowing to understand the features relevant to obtain high-performance in this bioinformatics domain. To the best of our knowledge, this study is the first to perform such evaluation for epistasis detection. The proposed approaches are able to surpass the performance of state-of-the-art works in the tested platforms, achieving an average speedup of 3.9$\times$ (7.3$\times$ on CPUs and 2.8$\times$ on GPUs) and maximum speedup of 10.6$\times$ on Intel UHD P630 GPU.

\end{abstract}


\section{Introduction}
\label{sec:intro}

During the last years, \ac{GWAS}~\cite{visscher201710} were able to show the relationship between the risk of an individual developing certain diseases with its genetic profile, through exploratory genetic analysis on data sets with groups of patients. With this information, the identified gene combination can be inspected by healthcare services, contributing to early diagnosis, prevention and provision of medical treatments personalized to each patient in order to reduce the risk of disease development~\cite{nature}. Hence, understanding the genetic causes of such conditions can offer invaluable insight into future treatments, as well as to serve as a crucial part of early disease diagnostic and prevention. Genetic studies usually focus on \acp{SNP}, \textit{i.e.}, gene variations that occur in a nucleotide on a given position of a DNA sequence, which can interact with each other to cause a disease, in a process called epistasis. While some diseases, such as Crohn's disease, are related to an interaction between two \acp{SNP}, \textit{i.e.}, second-order epistasis~\cite{dinu2012snp}, others can only be identified when considering high-order epistasis, \textit{i.e.}, interactions of three or more \acp{SNP}~\cite{sun2014hidden,yang2010interactions}. 

To accurately identify the \ac{SNP} interactions, it is necessary to exhaustively evaluate all the genetic combinations in a given biological data set. As the number of genetic interactions grows exponentially with the number of tested genes, this task becomes highly computationally demanding in modern computing systems, in particular when considering higher interaction orders. Evolutionary algorithms~\cite{gonccalves2020parallel} or machine learning methods~\cite{jiang2009random,shang2019review} can be used to speed up epistasis detection by narrowing down the search space. These approaches may lead, however, to reduced accuracy in identifying epistasis. Thus, the use of exhaustive search methods is the only way to guarantee the identification of the most accurate solution. As such, achieving high performance execution of exhaustive search epistasis methods on modern computing systems is particularly important, and a crucial step in tackling the treatment and prevention of diseases in future healthcare options.

State-of-the-art approaches for exhaustive search of genetic interactions are typically deployed on \acp{CPU}~\cite{goudey2015high} and \acp{GPU}~\cite{yung2011gboost,wienbrandt20191000}, although several works also target accelerators, and \acp{FPGA}~\cite{gonzalez2015parallel,wienbrandt2014fpga}. Regarding \acp{CPU} and \acp{GPU}, there is a considerable variety of devices available from different vendors (\textit{e.g.}, Intel, AMD and NVIDIA), each with distinct capabilities and architectures. Given the broad optimization space that arises from resource diversity, correlating the characteristics of the underlying hardware and the execution behavior of epistasis detection becomes a challenge, which limits the ability to select the most suitable devices for high-performance execution. As this issue is not unique to epistasis detection, but also common to other bioinformatics applications that share similar specifics to the workload here considered, such  as  the  use  of  population  count instructions~\cite{10.1145/2389241.2389247, lacour2015novel, 10.1093/bioinformatics/btq057}, the insights and the optimization/evaluation methodology herein proposed can also be applied to those applications, contributing to speed up the execution of several methods that evaluate genetic data. This can result in important implications on early disease detection and prevention, and in the well-being of the patients.

To tackle these issues, this work performs an exhaustive study on modern \ac{CPU} and \ac{GPU} architectures from different manufacturers, by proposing as case study several approaches to perform identification of three-way gene interactions. Each of the three-way epistasis detection approaches considered in this work are developed to exploit distinct hardware components in-built in current systems, which allows identifying specific micro-architectural features that are most relevant for attaining high-performance. This task is supported by the characterization of these approaches with \ac{CARM}~\cite{6506838}. The experimental evaluation is performed on 5 \acp{CPU} and 8 \acp{GPU} from different manufacturers, including state-of-the-art Intel Ice Lake SP and AMD Zen2 \acp{CPU}, and NVIDIA Ampere, AMD RDNA2 and Intel Xe \acp{GPU}, as well as previous architectures from each vendor. By considering past architectures, it is also possible to assess which micro-architectural features introduced through different generations benefit the identification of three-way gene interactions. To the best of our knowledge, this is the first attempt at tackling epistasis detection in all modern platforms, by also focusing on performance portability to efficiently explore the hardware resources offered by all main vendors. The proposed approaches are able to surpass the performance of state-of-the-art works in all platforms tested, achieving performance gains up to 10.6$\times$. This work features the following contributions:

\begin{itemize}
    \item \ac{CPU} and \ac{GPU} approaches\footnote{Available on (link to github repository)} for third-order exhaustive epistasis detection, that exercise different hardware components in each computing device, which can be used to relate the capabilities of the underlying hardware and the specifics of the application; 
    \item characterization of the proposed \ac{CPU} and \ac{GPU} approaches in \ac{CARM}, demonstrating how the utilization of different hardware components affects the execution of the applications, as well as to demonstrate the ability of the proposed methods to fully exploit all the capabilities of each type of device;
    \item extensive three-way gene interactions study in state-of-the-art \ac{CPU} and \ac{GPU} architectures from three different vendors (Intel, NVIDIA and AMD), in order to decouple the micro-architectural components more relevant for a high-performance execution of exhaustive epistasis detection and bioinformatics applications in general.
\end{itemize}

This article is structured as follows. Section \ref{sec:related} contains the analysis of the state-of-the-art works in epistasis detection. Section \ref{sec:problem} introduces the problem formulation of the epistasis detection. In Section \ref{sec:proposal}, the approaches considered for third-order exhaustive epistasis detection in this work are detailed. Section \ref{sec:results} presents the results of experimental evaluation. Section \ref{sec:conc} concludes the work and presents future directions.

\section{Related Work}
\label{sec:related}

Several works in the state-of-the-art explore two-way and three-way epistasis detection, focusing mostly on targeting either multi-core \acp{CPU}~\cite{6968770, goudey2015high} or \acp{GPU}~\cite{goudey2015high, joubert2018attacking, ipdps, tpds}. This article proposes several optimized approaches for exhaustive epistasis detection in both \ac{CPU} and \ac{GPU}, and presents an assessment of the effects of improvements in architecture across several devices and platforms. Some works feature approaches for epistasis detection without exhaustive search~\cite{sun2014analysis, guo2019epi, jiang2009random, shang2019review}, as exhaustive search methods have a high computational complexity. These approaches offer higher performance at the cost of lower result accuracy.

The majority of works target second-order epistasis, with epiSNP~\cite{7345653}, GBOOST~\cite{yung2011gboost}, multiEpistSearch~\cite{doi:10.1177/1094342015585846} and GWIS$_{FI}$~\cite{wang2014gwis} being examples. As higher epistasis orders have been linked to complex diseases, it is of great importance to propose high-performance approaches for this problem. MPI3SNP~\cite{doi:10.1177/1094342019852128} is a reference work for third-order exhaustive epistasis detection, targeting multi-\ac{CPU}/\ac{GPU} clusters. While it uses a binary format to reduce data size, and bitwise logic operations, the work presented in this article proposes performance optimizations that allow to fully exploit the effectiveness of memory accesses on both \ac{CPU} and \ac{GPU}, using cache blocking and achieving coalesced memory accesses.

Due to the data-parallel nature of this application, various relevant works focusing on identifying three-way gene interactions use \acp{GPU}. These include~\cite{gonzalez2015gpu}, which uses NVIDIA \acp{GPU}, and~\cite{jsspp}, which uses NVIDIA \acp{GPU} alongside a \ac{CPU}. However, such works focus on devices from a single vendor, which does not guarantee their applicability to other \acp{GPU} architectures. The approaches herein proposed also focus on portability in order to make use of hardware that is widely available in current computing devices. This allows to deploy the developed applications in platforms with multiple \ac{CPU} and \ac{GPU}, providing the means to select the most suitable device to perform exhaustive three-way epistasis.

A different approach is taken in works~\cite{ipdps} and~\cite{tpds}, that target NVIDIA \acp{GPU} for third order exhaustive epistasis detection, but specifically make use of the tensor cores available in newer NVIDIA architectures. However, to target these specific units, it is necessary to rely on vendor specific programming frameworks and constructs (\textit{e.g.} CUDA for NVIDIA), restricting the application deployment to specific \acp{GPU}. To provide portability across different devices and device types from different vendors, the approaches herein proposed are deployed using DPC++, which currently does not provide support for tensor cores. Although the proposed approaches do not directly exploit the features of this specialized hardware, they are devised to efficiently exploit the compute capabilities of current and upcoming \ac{CPU}, and \ac{GPU} platforms.

The work in~\cite{10.1007/978-3-030-57675-2_38} uses \ac{CPU} and \ac{GPU} in a collaborative way, performing two-way and three-way epistasis detection using \ac{CPU} and \ac{GPU}. However, the proposed framework focuses on the architectures of Intel \acp{CPU} and integrated \acp{GPU}. To the best of our knowledge, the work proposed in this manuscript is the first to feature experimental evaluation of highly-optimized approaches for exhaustive three-way gene interactions for \ac{CPU} and \ac{GPU} architectures on devices from all main vendors. This  allows to identify the main hardware features present in current computing systems that result in an efficient execution, enabling the selection of the most suitable devices to speedup the evaluation of three-way gene interactions, leading to improvements in the performance of bioinformatics applications and consequently the development of future techniques of disease diagnostic. 

\section{Problem Formulation}
\label{sec:problem}

Epistasis detection aims to relate the occurrence of phenotypic characteristics in individuals to interactions between \acp{SNP}. To do so, it relies on data sets containing the genetic information of a set of \acp{SNP} and patients/samples contained in a case-control data set D, with size $N \times (M+1)$, where N is the number of samples and M is the number of \acp{SNP}. Each entry D[i,j] in the data set corresponds to the genotype value of the i-th \ac{SNP} for the j-th sample, with $i \in 1, ...,M$ and $j \in 1, ...,N$. Entries D[M+1,j] correspond to the phenotype value, or disease state, for the i-th sample. The genotype can take values 0 (homozygous major allele), 1 (heterozygous allele) or 2 (homozygous minor allele), and the phenotype values can be either 0 (control) or 1 (case).

\begin{figure}[t]
    \centering
	\includegraphics[width=0.98\linewidth]{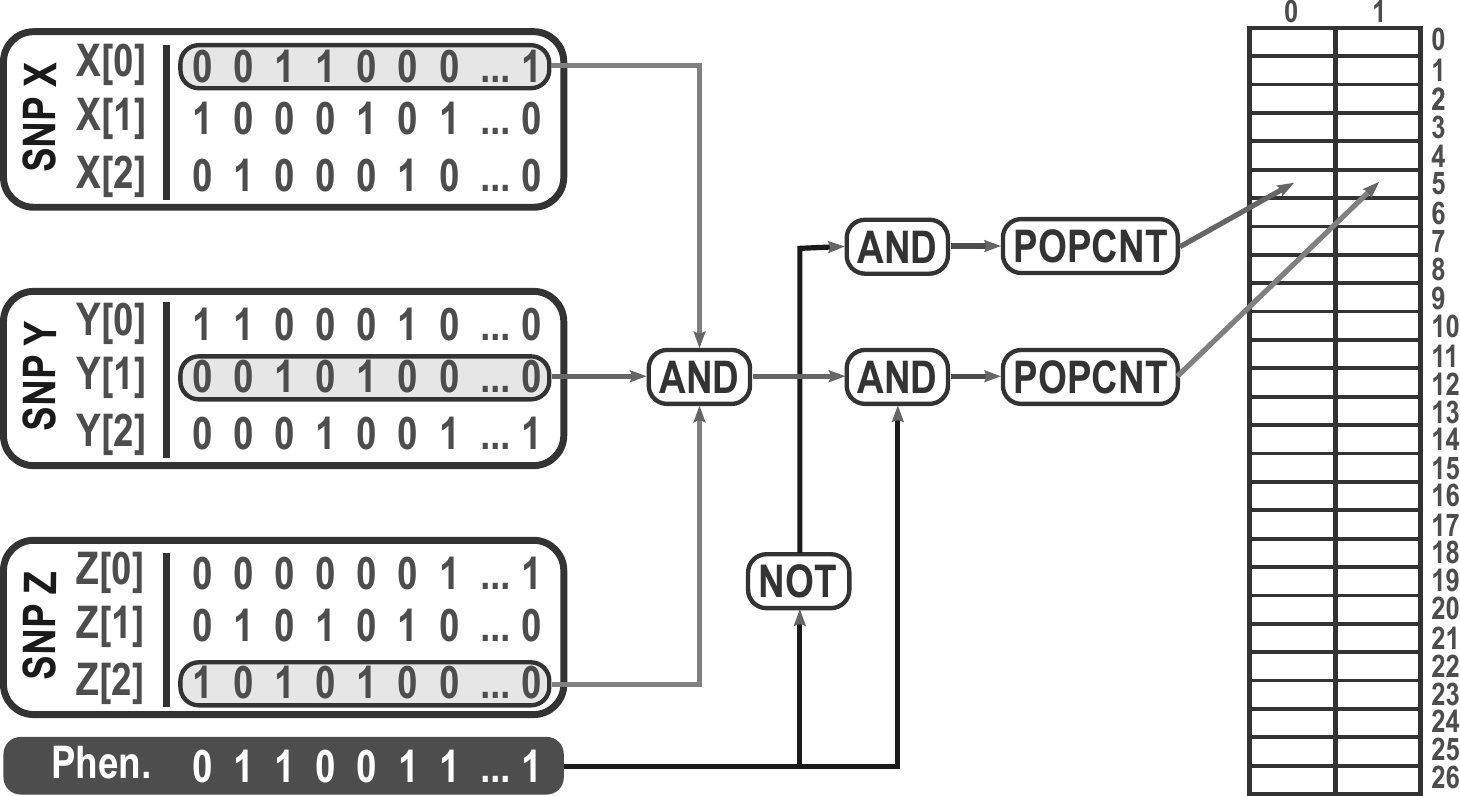}
    \caption{Binarized data set and frequency table construction.}
    \label{fig:data_bin}
\end{figure}

As the data sets containing the genetic information can have a large amount of samples and \acp{SNP}, it is crucial to reduce their memory requirements in order to achieve better performance. For this reason, this work considers the binarized format proposed in~\cite{wan2010boost}. As it can be observed in Figure~\ref{fig:data_bin}, a given \ac{SNP} X is represented by three arrays (X[0], X[1] and X[2]), on per each possible value of the genotype. The information for each sample is encoded in a single bit in each component, depending on the correspondent genotype. The phenotype is also represented in a binary format, with each bit indicating the phenotype for a given sample. 
 
Furthermore, the analysis of the genetic interaction on the D[i,j] is supported by a frequency table that stores the occurrence of each genotypic combination across all case and control samples. As the genotype can assume three distinct values (0, 1 or 2), for three-way epistasis detection this frequency table has a size of $3^3 = 27$ rows, and 2 columns, one for controls and one for cases. As exemplified in Figure~\ref{fig:data_bin}, the na{\"i}ve approach to fill the cells that correspond to the genotypic combination $(X,Y,Z)=(0,1,2)$, an \textit{AND} operation is done with the components $X[0]$, $Y[1]$ and $Z[2]$, followed by an \textit{AND} with the negated phenotype (controls) or by an \textit{AND} with the phenotype (cases). Finally, a \textit{POPCNT} operation is applied to both outputs with the result being added to the frequency table. This process is performed for all possible genotype/phenotype combinations for the considered \acp{SNP}, and is repeated until the information on all patients has been obtained.

Following this process, the information gathered in the frequency table is used to evaluate the \ac{SNP} combination. This is done by using objective functions that rely on biological and statistical criteria, in order to measure the likelihood of a gene combination being epistatic. This work considers the Bayesian K2 Score objective function~\cite{K2Score}, which is based on Bayesian network principles. For a given combination of $k$ \acp{SNP}, the K2 Score is given by
\begin{equation}
K2 = \sum_{i=1}^{I}\left(\sum_{b=1}^{r_{i}+1}\log (b)-\sum_{j=1}^{J}\sum_{d=1}^{r_{ij}}\log (d)\right),
\label{eq:1}
\end{equation}
with $I$ being the number of possible genotypic combinations among $k$ \acp{SNP} ($I = 3^k$), $J$ the number of diseases status ($J=2$ in case-control scenarios), $r_{i}$ the frequency of a certain genotypic combination $i$ at the evaluated \acp{SNP} $x=[x_1, x_2, ..., x_k]$, and $r_{ij}$ the number of samples that satisfy the occurrence of the phenotypic state $j$ with the genotypic combination $i$ at the evaluated \acp{SNP}.
The solution to the problem is the \ac{SNP} combination with the lowest K2 Score.

\section{Three-way Gene Interaction Detection Approaches for Multiple Execution Scenarios}
\label{sec:proposal}

In order to relate the characteristics of the underlying hardware and the specifics of the algorithm to perform evaluation of three-way gene interactions, this work considers four different methods to perform this task. Each method is expected to exercise different components of \acp{CPU} and \acp{GPU}, and the utilization of profiling tools and insightful modeling allows to assess which one is more suitable to achieve high performance in current hardware. In this work, all approaches use 32-bit integers to compress the input data set, due to their compatibility with all the considered devices/architectures.

\subsection{CPU Approaches}

As the biological data sets considered for the evaluation of gene interactions can contain thousands of samples and \acp{SNP}, the na{\"i}ve approach presented in Figure~\ref{fig:data_bin} is expected to be completely memory bound by the bandwidth of the slower memory levels, \textit{i.e.} LLC and \ac{DRAM}. Hence, the only possibility to achieve high-performance execution with this approach would be to drastically increase the capacity and bandwidth of L1 and/or L2 caches, a configuration not present in current hardware.

One solution for this issue is to reduce the memory requirements of the algorithm. Since the genotype can have at most three values, the component of \ac{SNP} X correspondent to the genotype 2 (X[2]) can be inferred from X[0] and X[1] through a \textit{NOR} operation. The data set can also be divided in two components, one containing all control samples, and other containing all cases. This way, each \ac{SNP} is represented by four arrays, two per each phenotype, and the phenotype array is not required in the frequency table construction for the evaluation of each gene combination, further reducing the computational requirements of the application. With these optimizations, the amount of memory transfers is expected to reduce by 1/3. However, while the first method would require a total of $27 \times 6 = 162$ compute instructions to evaluate all the genotype combinations for each processed element, by removing the phenotype, this task only requires a total of $(3$ \textit{NOR} $+ 1$ \textit{AND} $+ 1$ \textit{POPCNT} $)\times27 = 57$, \textit{i.e.}, the amount of computations performed will reduce around 65\%. This hints to a reduction of the \ac{AI} of the application, indicating the possibility for the kernel to become more memory bound. 

In this scenario, a third method to perform epistasis detection can include loop tiling techniques to improve the data locality, maximizing the cache utilization. Since cache blocking techniques do not affect the amount of memory transfers and performed computations, the \ac{AI} is not expected to change and the performance of the kernel will likely improve. As observed in Algorithm~\ref{alg:cpu}, in this method to perform three-way epistasis detection, each \ac{CPU} core tackles simultaneously three blocks with $B_S$ \acp{SNP} and $B_P$ samples. Each block $B_S$ is controlled by the loop variables $i_{0}, i_{1}$ and $i_{2}$, while the variables $ii_0, ii_1$ and $ii_2$ iterate over the \acp{SNP} contained in each block. Similarly, the variable $p_0$ controls the block $B_P$ and the loop $p$ iterates over each sample. With this approach, up to $B_S^3$ combinations can be processed simultaneously, thus requiring a frequency table array of size $B_S^3 \times 2 \times 27$. Per each valid combination, the genotypes 0 and 1 of each \ac{SNP} correspondent to controls ($X_0[0], X_0[1]$) and cases ($X_1[0], X_1[1]$) are loaded from memory, making a total of six loads per evaluated combination. The components of the genotype 2 are obtained with the three \textit{NOR} operations, one per each \ac{SNP}. The next step is the evaluation of the gene interaction for all the possible genotype combinations through the \textit{AND} and \textit{POPCNT} instructions, and the output is used to update the frequency table $ft_{0|1}$. After all samples are evaluated, the score is obtained from the get\_score function. To parallelize this algorithm, each core fetches a task from a thread pool. Each thread performs a set of combinations, which can be defined dynamically in order to improve load balancing. To avoid synchronization barriers between tasks, the scores are kept locally to each thread and a final reduction is performed to obtain the global solution.

\SetInd{0.4em}{0.4em}
\begin{algorithm}[t]
\small
\caption{Epistasis detection on each CPU core.}
\label{alg:cpu}
\KwData{$D_{0|1}$}
\KwResult{$score$}
\For{$i_0,i_1,i_2 \gets 1$ \KwTo $M/B_S$}{
    $ft_{0|1} \gets 0$\;
    \For{$p_0 \gets 1$ \KwTo $N_{0|1}/B_P$}{
        \For{$ii_0,ii_1,ii_2 \gets 1$ \KwTo $B_S$}{
            \If{$ii_2 > ii_1 > ii_0$}{
                \For{$p \gets 1$ \KwTo $B_P$}{
                    $X_{0|1}(0) \gets $LOAD$(D_{0|1}[i_0,ii_0,p_0,p,0])$\;
                    $X_{0|1}(1) \gets $LOAD$(D_{0|1}[i_0,ii_0,p_0,p,1])$\;
                    $Y_{0|1}(0) \gets $LOAD$(D_{0|1}[i_1,ii_1,p_0,p,0])$\;
                    $Y_{0|1}(1) \gets $LOAD$(D_{0|1}[i_1,ii_1,p_0,p,1])$\;
                    $Z_{0|1}(0) \gets $LOAD$(D_{0|1}[i_2,ii_2,p_0,p,0])$\;
                    $Z_{0|1}(1) \gets $LOAD$(D_{0|1}[i_2,ii_2,p_0,p,1])$\;
                    $X_{0|1}(2) \gets $NOR$(X_{0|1}(0),X_{0|1}(1))$\;
                    $Y_{0|1}(2) \gets $NOR$(Y_{0|1}(0),Y_{0|1}(1))$\;
                    $Z_{0|1}(2) \gets $NOR$(Z_{0|1}(0),Z_{0|1}(1))$\;
                    \For{$g_X,g_Y,g_Z \gets 0$ \KwTo $2$}{
                        $ymm_{0|1}(g_X,g_Y,g_Z) \gets$AND$(X_{0|1}(g_X),$
                        
                        $Y_{0|1}(g_Y), Z_{0|1}(g_Z))$\;
                        
                        $ft_{0|1}(g_X,g_Y,g_Z)\gets ft_{0|1}(g_X,g_Y,g_Z)+$
                        
                        POPCNT$(ymm_{0|1}(g_X,g_Y,g_Z))$\;
                    }
                }
            }
        }
    }
    $score \gets$get\_score$(ft_{0|1})$\;
}
\end{algorithm}

The parameters $B_S$ and $B_P$ used for loop tiling are determined according to the capacity of the L1 data cache, by defining a maximum size for the frequency table ($size_{FT}$) and for each block $B_S \times B_P$ ($size_{Block}$), such that both these components fit in the first cache level. As the frequency table occupies $2 \times 27 \times B_S^3 \times \beta_{int}$ bytes, with $\beta_{int} = 4$ B for 32-bit integers, $B_S$ can be determined through the equation $B_S^3 \times \beta_{int} \times 2 \times 27 \leq size_{FT}$. After defining $B_S$, $B_P$ can be calculated through the equation $B_S \times B_P \times \beta_{int} \times 2 \leq size_{Block}$. With this, the optimal values for $B_S$ and $B_P$ can be determined for each \ac{CPU}, allowing for this implementation to achieve efficient cache blocking in the different \acp{CPU}. For example, for an Intel Ice Lake SP \acp{CPU} with a L1 data cache of 48 kB, if seven ways are used for the frequency table ($size_{FT} = 28$ kB) and four ways for the block ($size_{Block} = 16$ kB), then $B_S \leq 5.1$ and $B_P \leq 409.6$.

\begin{algorithm}[t]
\caption{Epistasis detection on GPU.}
\small
\label{alg:gpu}
\KwData{$D_{0|1}$}
\KwResult{$score$}
$(i_0,i_1,i_2) \gets$thread\_id$(0,1,2)$\;
\If{$i_2 > i_1 > i_0$}{
    $ft \gets 0$\;
    \For{$p \gets 1$ \KwTo $B_S \times N_{0|1}/B_S$}{
        $X_{0|1}(0) \gets D_{0|1}[i_0,p,0]$\;
        $X_{0|1}(1) \gets D_{0|1}[i_0,p,1]$\;
        $Y_{0|1}(0) \gets D_{0|1}[i_1,p,0]$\;
        $Y_{0|1}(1) \gets D_{0|1}[i_1,p,1]$\;
        $Z_{0|1}(0) \gets D_{0|1}[i_2,p,0]$\;
        $Z_{0|1}(1) \gets D_{0|1}[i_2,p,1]$\;
        $X_{0|1}(2) \gets \sim(X_{0|1}(0) | X_{0|1}(1))$\;
        $Y_{0|1}(2) \gets \sim(Y_{0|1}(0) | Y_{0|1}(1))$\;
        $Z_{0|1}(2) \gets \sim(Z_{0|1}(0) | Z_{0|1}(1))$\;
        \For{$g_X,g_Y,g_Z \gets 0$ \KwTo $2$}{
            
            $ft(g_X,g_Y,g_Z)\gets ft(g_X,g_Y,g_Z)+$
            
            POPCNT$(X_{0|1}(g_X) \& Y_{0|1}(g_Y) \& Z_{0|1}(g_Z))$\;
        }
    }
    $score \gets$get\_score$(ft)$\;
}
\end{algorithm}

After introducing cache blocking, the application performance is likely to become limited by the compute capabilities. 
Thus, the fourth (and final) approach considered herein focuses on code vectorization, through the use of vector intrinsics for the \textit{LOAD}, \textit{NOR}, \textit{AND} and \textit{POPCNT} instructions in Algorithm~\ref{alg:cpu}. As different \ac{CPU} architectures are used in the scope of this work, different vectorization strategies are necessary depending on the target \ac{CPU}. In the case of \acp{CPU} that support at most AVX instructions, such as AMD Zen and Zen2 and Intel Kaby Lake architectures, the \textit{LOAD} and \textit{AND} operations are performed through the intrinsics \textit{\_mm256\_loadu\_si256} \textit{\_mm256\_and\_si256}, while for processors that support AVX512 instructions, the intrinsics \textit{\_mm512\_loadu\_si512} \textit{\_mm512\_and\_si512} are used instead. Furthermore, as none of the AMD and Intel processors support vectorized \textit{NOR}, this instruction is vectorized through the use of a  vector \textit{OR} (\textit{\_mm256\_or\_si256} or \textit{\_mm512\_or\_si512}) followed by a vector \textit{XOR} (\textit{\_mm256\_xor\_si256} or \textit{\_mm512\_xor\_si512}) with a vector with all bits set to one. Regarding the \textit{POPCNT} instructions, with the exception of Intel Ice Lake SP, the \acp{CPU} only support scalar \textit{POPCNT}. Thus, to update the frequency table, this operation is applied to each element of the vector, which is extracted with the instructions\textit{\_mm256\_extract\_epi64} for AVX, or \textit{\_mm256\_extract\_epi64} and \textit{\_mm512\_extracti64x4\_epi64} for AVX512. For Intel Ice Lake SP, as it supports AVX512 vector \textit{POPCNT} (\textit{\_mm512\_popcnt\_epi32}), each position of the frequency table is updated by applying a reduction operation (\textit{\_mm512\_reduce\_add\_epi32}) to the \textit{POPCNT} outputs. 

\subsection{GPU Approaches}

Similarly to the \ac{CPU} approaches, the method to perform three-way gene interactions presented in Figure~\ref{fig:data_bin} is expected to be completely limited by the main memory of the \ac{GPU}. For this reason, the second approach considered for the \ac{GPU} follows the optimizations introduced in the \ac{CPU}, \textit{i.e.}, inferring genotype 2 from the other genotypes, and separating the data set according to the phenotype. As the \ac{AI} is expected to reduce, to further improve the performance of this method, additional memory-related optimizations are implemented.

\begin{table}[t]
\centering
\caption{CPU Devices used in the experimental evaluation.}
\label{tab:cpu_systems}
\resizebox{0.98\linewidth}{!}{%
\begin{tabular}{@{}ccccc@{}}
\toprule
System & \begin{tabular}[c]{@{}c@{}}CPU Device  \\ Arch. \textpipe~Base Freq. [GHz]\end{tabular}                   & Cores &  \begin{tabular}[c]{@{}c@{}} Vector \\ Width \end{tabular} (ISA) \\ \midrule
CI1    & \begin{tabular}[c]{@{}c@{}}Intel\textregistered~Core\texttrademark~i7-8700K  \\ SKL \textpipe~3.7\end{tabular}             & 6 & 256-bit (AVX)                                                                  \\
CI2    & \begin{tabular}[c]{@{}c@{}}(2x) Intel\textregistered~Xeon\textregistered~Gold 6140   \\ SKX \textpipe~2.3\end{tabular}       & (2x)18 & 512-bit (AVX512)                                                                \\
CI3    & \begin{tabular}[c]{@{}c@{}}(2x) Intel\textregistered~Xeon\textregistered~Platinum 8360Y  \\ ICX \textpipe~2.4\end{tabular}  & (2x)36 & 512-bit  (AVX512)                                                              \\
CA1    & \begin{tabular}[c]{@{}c@{}}AMD EPYC\texttrademark~7601 \\ Zen \textpipe~2.2\end{tabular}                   & 64 & 128-bit (AVX)                                                                 \\
CA2    & \begin{tabular}[c]{@{}c@{}}AMD EPYC\texttrademark~7302P          \\ Zen2 \textpipe~3.0\end{tabular}     & 16 & 256-bit (AVX)                                                                    \\ \bottomrule
\end{tabular}%
}
\end{table}

Since the evaluations of \ac{SNP} combinations are independent from each other, the best way to parallelize the algorithm in a \ac{GPU} is to attribute each thread to a single combination. However, the input data presented in Figure~\ref{fig:data_bin} contains the \acp{SNP} placed in rows and the samples in columns, which prevents coalesced memory accesses, since each \ac{SNP} is separated in memory by N samples. Hence, the third method herein proposed considers a transposed data set, such that the \acp{SNP} are organized by columns and the consecutive samples are organized in rows. This increases the chances of consecutive threads using data that is stored consecutively in memory, leading to coalesced memory accesses loads instead of memory gather and scatter operations.

Despite this, in larger data sets, this approach will likely suffer from memory bottlenecks, as each sample is separated in memory by M \acp{SNP}. To mitigate this issue, the fourth \ac{GPU} procedure is tiled with the \acp{SNP} organized in blocks of size $B_S$, by placing $B_S$ \ac{SNP} values from the same sample adjacently. This allows to improve memory accesses and cache utilization. With this organization, and by defining the size of each group of threads to be $B_S$, the chances of achieving coalesced memory accesses increase, by maintaining an interval of $B_S$ between consecutive samples for the same \ac{SNP}. The optimal value of $B_S$ depends on the target \ac{GPU}, but for most architectures is defined as a multiple of 32 or 64.

The \ac{GPU} kernel correspondent to the fourth \ac{GPU} approach is represented in Algorithm~\ref{alg:gpu}. Each enqueueing of the \ac{GPU} kernel receives multiple blocks of $B_{Sched}^3$ combinations from the \ac{CPU}, processing them in succession until all combinations have been evaluated. The value of $B_{Sched}$ is defined empirically for each \ac{GPU} device, in order to maximize performance on each \ac{GPU} architecture. Each kernel execution contains $B_{Sched}^3$ threads, with each kernel instance processing one combination. As it can be observed in Algorithm~\ref{alg:gpu}, the assigned \ac{SNP} combination $(i_0,i_1,i_2)$ is obtained from the multi-dimensional thread index. The frequency table $ft$ is created and initialized in the private memory, eliminating the need for synchronization between different threads, and making use of the fast-access memory in the \ac{GPU} general register file. The frequency table is filled by going through all samples $p$ in the data set. Following the completion of the frequency table $ft$, the score for the combination is obtained with the function get\_score. Each \ac{GPU} thread compares the value of the current score to its current best score, and updates it if necessary. Following the processing of all combination blocks, the final solution is determined in the \ac{CPU} host from the best scores obtained by all \ac{GPU} threads.

\section{Experimental Results}
\label{sec:results}

The experimental evaluation herein performed aims at assessing the impact of micro-architectural improvements when executing third-order epistasis detection across different generations of computing systems. As first step, the \ac{CPU} and \ac{GPU} approaches (Section~\ref{sec:proposal}) are characterized in \ac{CARM}~\cite{6506838, advisor}, contained in Intel\textregistered~Advisor 2021.4, which is part of the Intel\textregistered~oneAPI toolkit, to identify the more suitable method to achieve high-performance execution.

\begin{table}[t]
\centering
\caption{GPU Devices used in the experimental evaluation (\textbf{*} values obtained experimentally).}
\label{tab:gpu_systems}
\resizebox{0.95\linewidth}{!}{%
\begin{tabular}{@{}ccccc@{}}
\toprule
System & \begin{tabular}[c]{@{}c@{}} GPU Device \\ Arch. \textpipe~Boost Freq. {[}GHz{]}\end{tabular} & \begin{tabular}[c]{@{}c@{}}Comp. \\ Units (CU)\end{tabular} & \begin{tabular}[c]{@{}c@{}}Stream \\ Cores \end{tabular}  & \begin{tabular}[c]{@{}c@{}}POPCNT \\ per CU.\end{tabular} \\ \midrule
GI1    & \begin{tabular}[c]{@{}c@{}} Intel\textregistered~Graphics UHD P630  \\  Gen9.5 \textpipe~1.200 \end{tabular}    & 24 & 192     & 4\textbf{*}                                                                 \\
GI2    & \begin{tabular}[c]{@{}c@{}} Intel\textregistered~Iris Xe MAX  \\  Gen12 \textpipe~1.650 \end{tabular}        & 96 & 768  & 4\textbf{*}                                                               \\
GN1    & \begin{tabular}[c]{@{}c@{}} NVIDIA Titan Xp  \\  Pascal \textpipe~1.582 \end{tabular}      & 30 & 3840    & 32                                                                 \\
GN2    & \begin{tabular}[c]{@{}c@{}} NVIDIA Titan V    \\  Volta \textpipe~1.455 \end{tabular}                & 80 & 5120    & 16                                                                   \\ 
GN3    & \begin{tabular}[c]{@{}c@{}} NVIDIA Titan RTX    \\  Turing \textpipe~1.770 \end{tabular}              & 72 & 4608    & 16                                                                   \\
GN4    & \begin{tabular}[c]{@{}c@{}} NVIDIA A100 (250W)    \\  Ampere \textpipe~1.410 \end{tabular}              & 108 & 6912    & 16                                                                   \\
GA1    & \begin{tabular}[c]{@{}c@{}} AMD Radeon\texttrademark~Pro VII  \\  Vega20 \textpipe~1.700 \end{tabular}      & 60 & 3840    & $\approx 12$\textbf{*}                                                                  \\
GA2    & \begin{tabular}[c]{@{}c@{}} AMD Instinct\texttrademark~Mi100  \\  CDNA \textpipe~1.502 \end{tabular}        & 120 & 7680    & $\approx 12$\textbf{*}                                                                  \\ 
GA3    & \begin{tabular}[c]{@{}c@{}} AMD Radeon\texttrademark~RX 6900 XT  \\  RDNA2 \textpipe~2.250 \end{tabular}                 & 80 & 5120    & $\approx 10$\textbf{*}                                                                  \\
\bottomrule
\end{tabular}%
}
\end{table}

\begin{figure*}[t]
    \centering
    \begin{subfigure}[t]{0.44\linewidth}
        \centering
	    \includegraphics[width=1\linewidth]{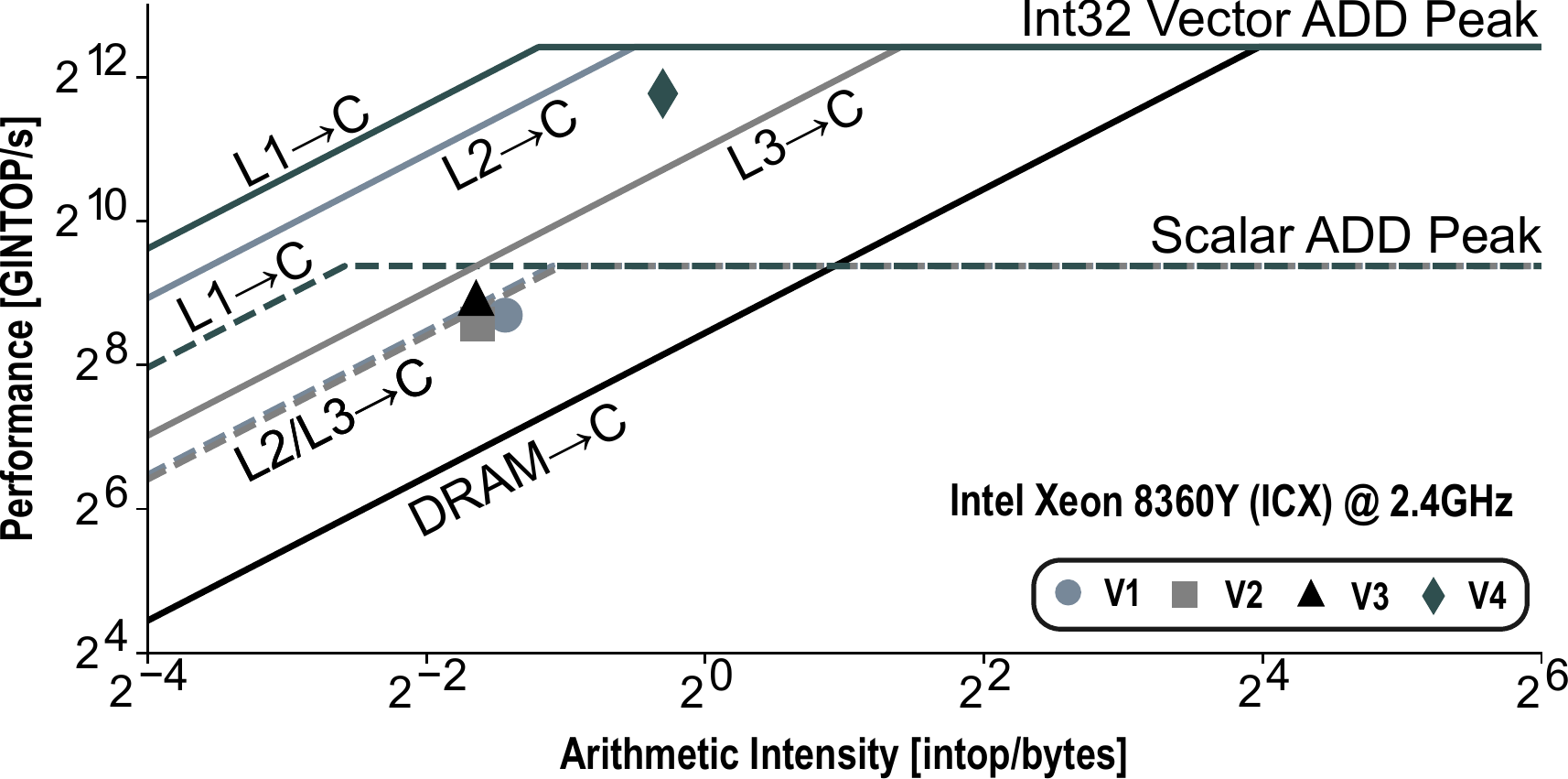}
	    \caption{Intel Xeon Platinum 8360Y (Ice Lake SP). Slashed roofs correspond to the scalar limits on the \ac{CPU}.}
	    \label{fig:cpu_roofline}
    \end{subfigure}%
    \begin{subfigure}[t]{0.44\linewidth}
        \centering
	    \includegraphics[width=1\linewidth]{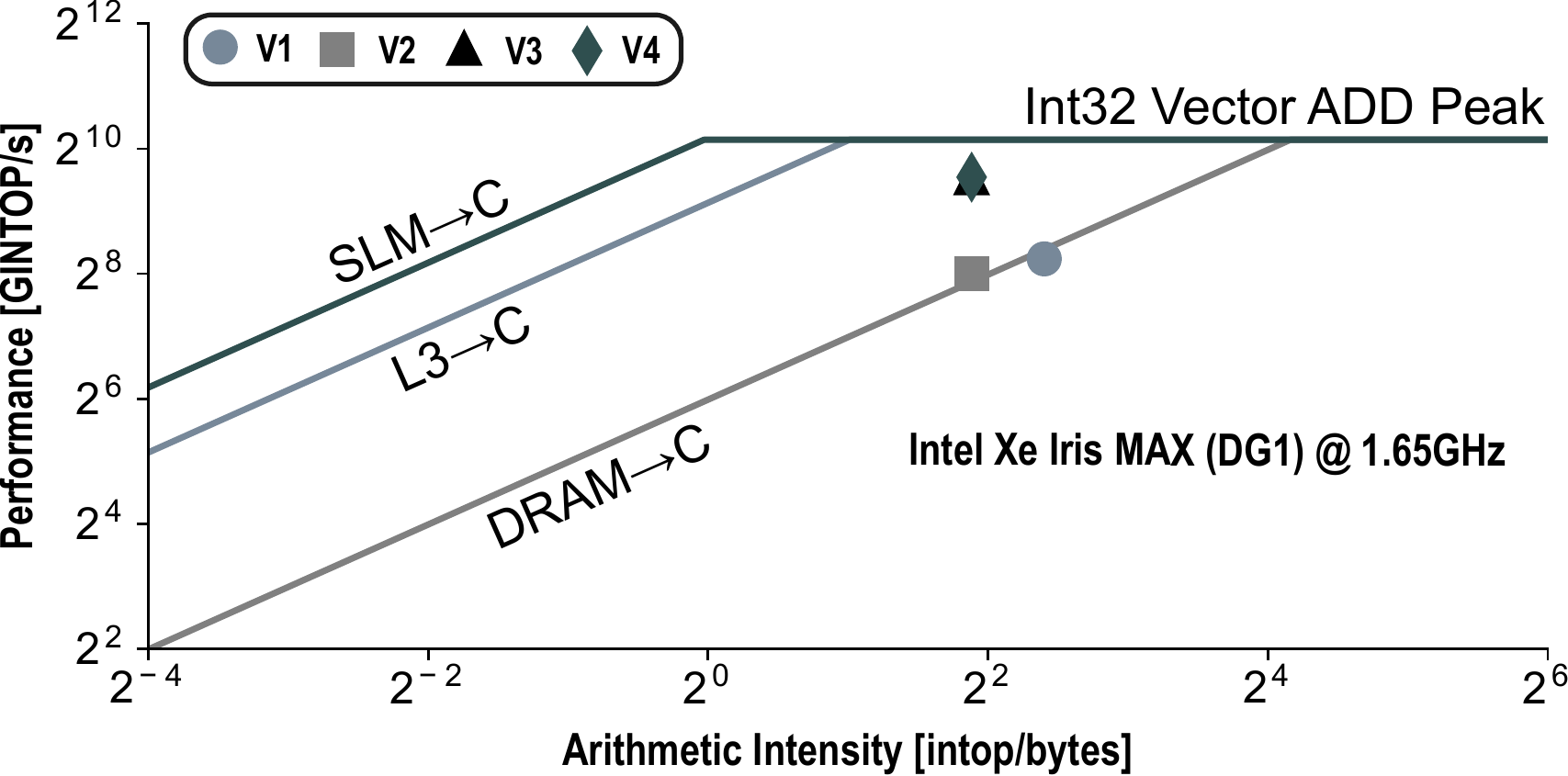}
    	    \caption{Intel Iris Xe Max (Gen12).}
	    \label{fig:gpu_roofline}
    \end{subfigure}%
    \caption{Evaluation of different aproaches on \ac{CARM} for latest Intel server \ac{CPU} and \ac{GPU}.}
    \label{fig:rooflines}
\end{figure*}

The most suitable version is then used to perform an extensive study on \acp{CPU} and \acp{GPU}, from Intel, AMD and NVIDIA, in order to decouple the most relevant micro-architectural features for the execution of three-way epistasis detection. The devices considered in this work are summarized in Tables~\ref{tab:cpu_systems} and~\ref{tab:gpu_systems}. In this work, NVIDIA multiprocessors, Intel execution units and AMD compute units are referred as compute units (CU), and CUDA cores, Intel \ac{GPU} SIMD4 instances and AMD stream cores are referred as stream cores. From Intel, we consider the Skylake (CI1), Skylake SP (CI2) and Ice Lake SP (CI3) micro-architectures. The AMD \acp{CPU} have the Zen (CA1) and Zen2 (CA2) micro-architectures. While the \acp{CPU} CI2 and CI3 support AVX512 instructions, CI1, CA1 and CA2 can at most perform AVX instructions. This work also considers two of the latest Intel GPU architectures, \textit{i.e.}, Gen9.5 (GI1) and Gen12 (GI2), as well as, the AMD Vega20 (GA1), CDNA (GA2) and RDNA2 (GA3), and NVIDIA Pascal (GN1), Volta (GN2), Turing (GN3) and Ampere (GN4). As observed in Table~\ref{tab:gpu_systems}, these \acp{GPU} have distinct \textit{POPCNT} capabilities, which is one of the main instructions in epistasis detection. GN1 has the highest throughput, of up to 32 \textit{POPCNT} per cycle and compute unit, while both Intel \acp{GPU} are only able to perform 4 \textit{POPCNT} per cycle and compute unit.

The \ac{GPU} approaches are implemented by using the  SYCL framework, \textit{i.e.}, the DPC++ programming model, to deploy the application in \acp{GPU} from any vendor while maintaining a high level of performance portability. DPC++ uses as back-ends CUDA 11.4 for NVIDIA, ROCm 4.2 for AMD, and Intel Level-Zero for Intel \acp{GPU}. Due to the lack of support for population count instructions in HIPSYCL, DPC++ does not guarantee a fair comparison between \acp{CPU} of different vendors. Thus, OpenMP programming model with dynamic scheduler is used instead, and the application is compiled with GCC-9.3 across all \ac{CPU} processors. The \ac{CPU} approaches are vectorized with vector intrinsics, which are transversal to any of the considered x86 processor from AMD and Intel. 

To compare devices with different core counts, frequencies and vector widths, the performance is scaled to the number of cores and vector width for \acp{CPU}, and to the number of computing units and stream cores for \acp{GPU}. To compare data sets with different dimensions, the performance is represented as the total number of elements processed per second or per cycle. The total number of elements is defined as the amount of processed combinations multiplied with the number of samples, \textit{i.e.}, $nCr(M, k) \times N$, where $nCr(M, k)$ is the number of $k$-combinations in a set of $M$ items, $M$ is the number of \acp{SNP}, $k$ the interaction order, and $N$ the number of individual samples. The experimental results are obtained for synthetic data sets equivalent to real case scenarios, containing \acp{SNP} ranging from 2048 to 8192 and 16384 samples.  

\begin{figure*}[t]
    \centering
    \begin{subfigure}[t]{0.32\textwidth}
        \centering
	    \includegraphics[width=1\textwidth]{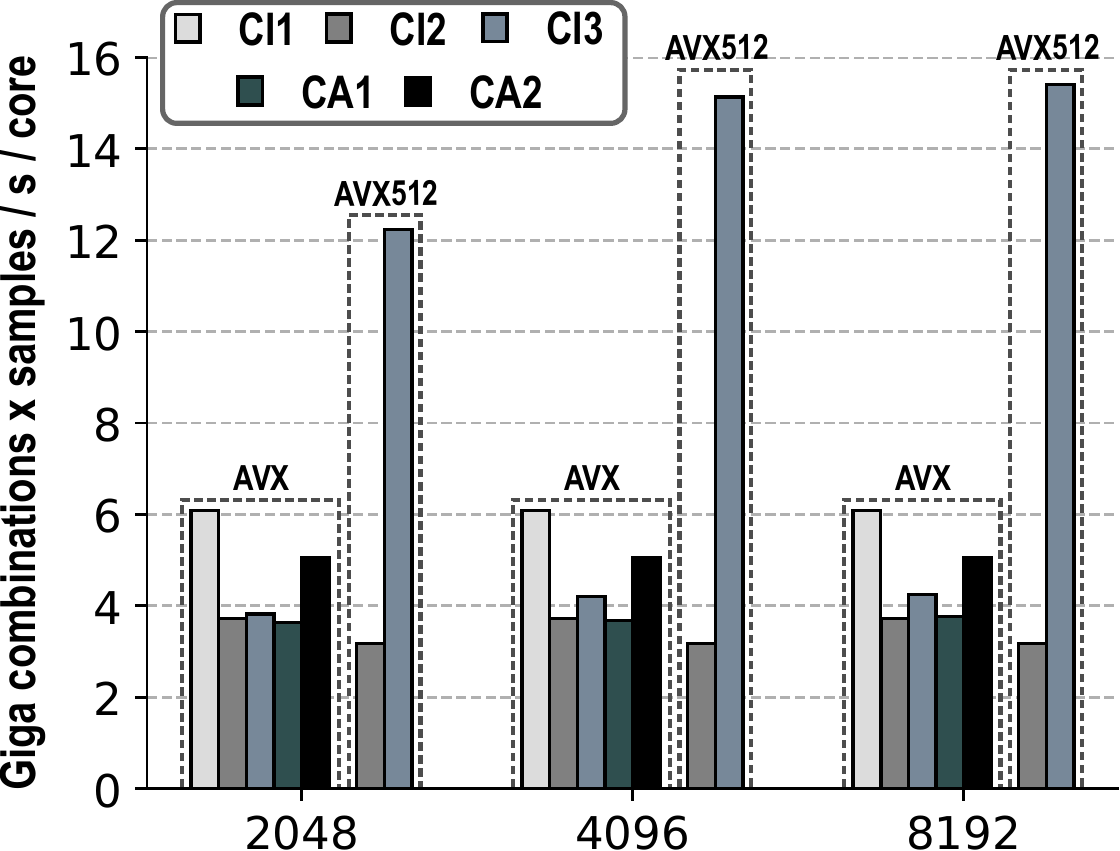}
	    \caption{Processed elements per second and \\ per core.}
	    \label{fig:cpu_sec_core}
    \end{subfigure}%
    \begin{subfigure}[t]{0.32\textwidth}
        \centering
	    \includegraphics[width=1\textwidth]{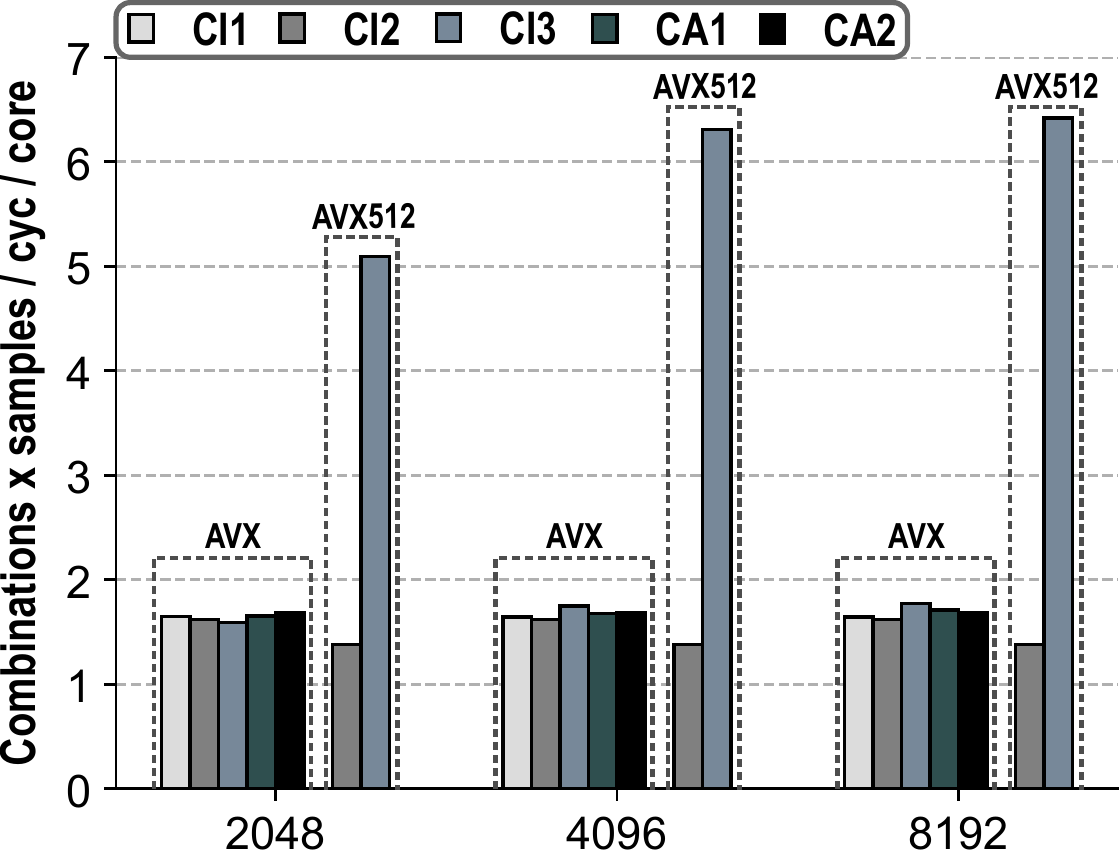}
	    \caption{Processed elements per cycle and \\ per core.}
	    \label{fig:cpu_cyc_core}
    \end{subfigure}%
    \begin{subfigure}[t]{0.32\textwidth}
        \centering
	    \includegraphics[width=1\textwidth]{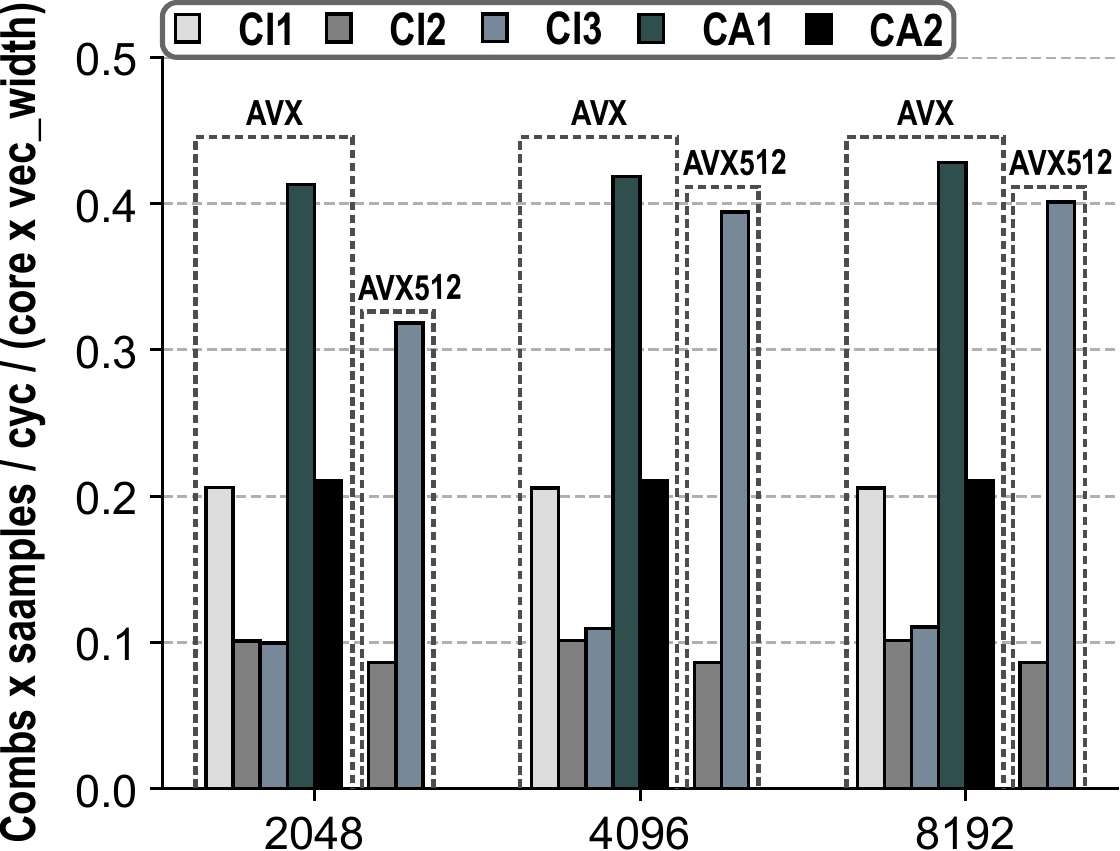}
	    \caption{Processed elements per cycle and per (core x vector width).}
	    \label{fig:cpu_cyc_core_vec}
    \end{subfigure}%
    \caption{\ac{CPU} performance evaluation for different number of \acp{SNP} (2048, 4096, 8192) and 16384 samples.}
    \label{fig:cpu_results}
\end{figure*}

\subsection{Characterization in Cache-Aware Roofline Model}

Figure~\ref{fig:rooflines} presents the characterization in \ac{CARM} of the different methods to perform three-way epistasis detection on the CI3 \ac{CPU} (Figure~\ref{fig:cpu_roofline}), and GI2 \ac{GPU} (Figure~\ref{fig:gpu_roofline}). For Intel Ice Lake SP (CI3), the baseline algorithm (V1) is limited by the memory bandwidth of one of the slower memory levels, in this case the scalar L3 bandwidth roof. By removing the third genotype and the phenotype (V2), as expected, this approach suffers from a reduction in the \ac{AI}, indicating a bigger impact of the memory and continues to be limited by the scalar L3 bandwidth roof. Although compared to the method V1, V2 achieves a speedup in the execution time around 2x, there is an apparent loss of performance due to a reduction of 2.1x on the amount of performed computations. Incorporating cache blocking in V3 resulted in an improvement in performance around 1.2x, moving the kernel to the top of L2 cache scalar roof and right below the scalar \textit{ADD} roof, indicating that the application is limited by the compute roof and private caches. By vectorizing the application (V4), there is an performance increase of 7.5x compared to V3, and the kernel is now bound by the integer vector \textit{ADD} peak. In comparison to the baseline algorithm, the performance increased 8.5x.

Similar insights can be obtained from the characterization of the \ac{GPU} approaches on \ac{CARM}. As shown in Figure~\ref{fig:gpu_roofline}, the na{\"i}ve procedure (V1) is completely memory bound by the \ac{DRAM} bandwidth. By inferring the genotype 2 from genotypes 0 and 1, and by splitting the data set in cases and controls (V2), there is a decrease in \ac{AI}, similar to the effect that occurs on \ac{CPU}. The performance also decreases, despite the improvement on the execution time around 1.79x when compared to V1. This is mainly due to the decrease of 47.5\% in the number of transferred bytes of memory being overshadowed by the decrease in total number of operations (2.11x). While V2 is still mainly limited by \ac{DRAM}, with the data set transposition, the coalesced memory accesses in the third method (V3) guarantee a significant performance increase, showing the efficacy of using the transposed data set. With the data set tiling introduced in the last \ac{GPU} method (V4), there is slightly improvement in performance, placing the application closer to the maximum performance of 32-bit integer vector instructions.

From this analysis, it is possible to conclude that to fully exploit current \acp{CPU} and \acp{GPU}, the fourth approach of each device is the most suitable to achieve high-performance execution, which will be used to experimentally evaluate the capabilities of the different \ac{CPU} and \ac{GPU} micro-architectures. Although the score calculation is included in the kernel evaluation of each considered approach, its contribution for the total execution time is residual, and only around 4\%, according to Intel\textregistered~Advisor. Thus, the performance represented in \ac{CARM} corresponds mostly to the frequency table construction, \textit{i.e.}, the most computational demanding part of epistasis detection.

\subsection{CPU Evaluation}

The performance results obtained for different data sets on diverse \acp{CPU} from Intel and AMD are presented in Figure~\ref{fig:cpu_results}. To perform a fair comparison between Intel processors that support AVX512 and the remaining systems, the AVX version of the algorithm used on the other processors is also executed on CI2 and CI3. The tiling parameters \textless$B_S, B_P$\textgreater~used in the \ac{CPU} approach are calculated by defining the size of the frequency table to fit in 7 ways of the L1 data cache in all \acp{CPU}. Since Ice Lake SP has a larger L1 data cache with 12 ways, the size of each block $B_S \times B_P$ is defined to occupy 4 ways in CI3, leaving one way for the prefetcher to exploit. Since the other systems only have a total of 8 ways, the size of each block is defined to occupy the remaining way. Thus, the experimental results on the \ac{CPU} are obtained with the configuration \textless$5, 400$\textgreater~for CI3, and \textless$5, 96$\textgreater~for the remaining \acp{CPU}. $B_P$ is rounded to the closest multiple of the number of 32-bit integers that fit in the vector registers.

\begin{figure*}[t]
    \centering
    \begin{subfigure}[t]{0.32\textwidth}
        \centering
	    \includegraphics[width=1\textwidth]{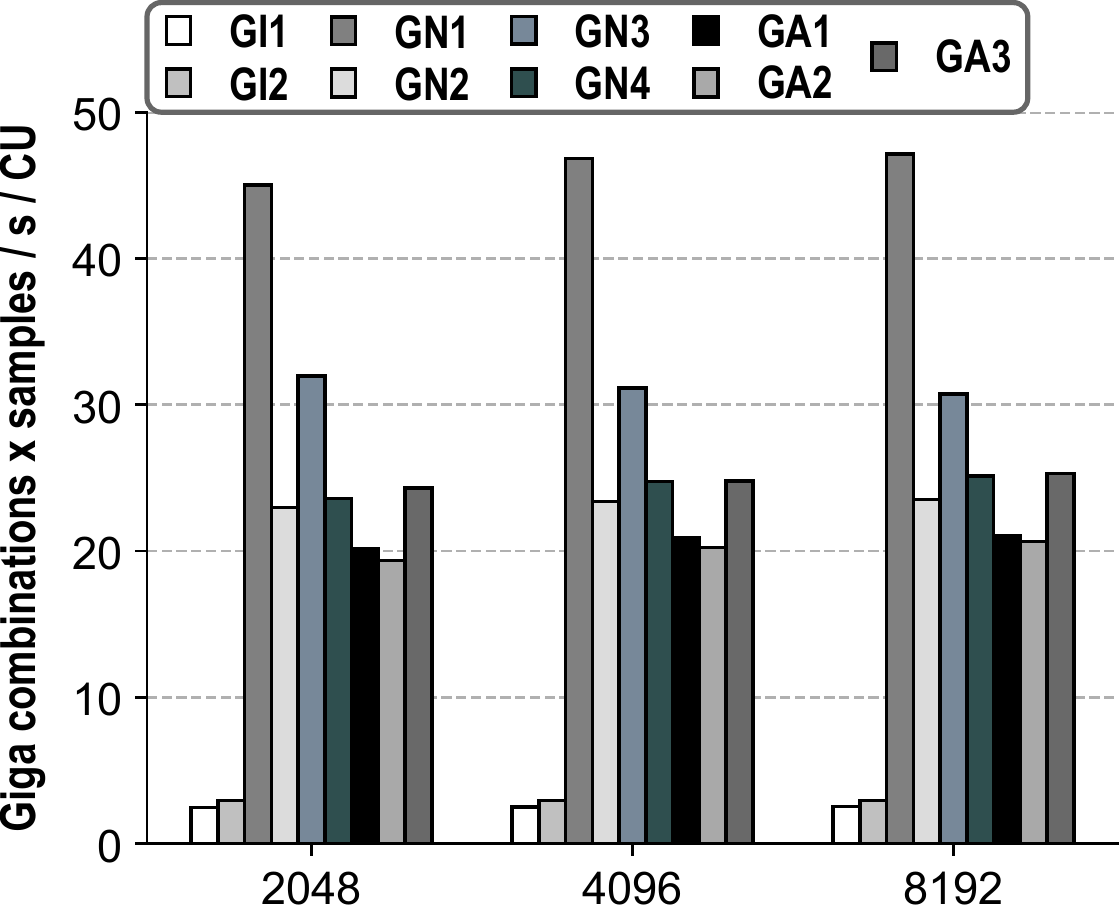}
	    \caption{Processed elements per second and \\ per compute unit.}
	    \label{fig:gpu_sec_core}
    \end{subfigure}%
    \begin{subfigure}[t]{0.32\textwidth}
        \centering
	    \includegraphics[width=1\textwidth]{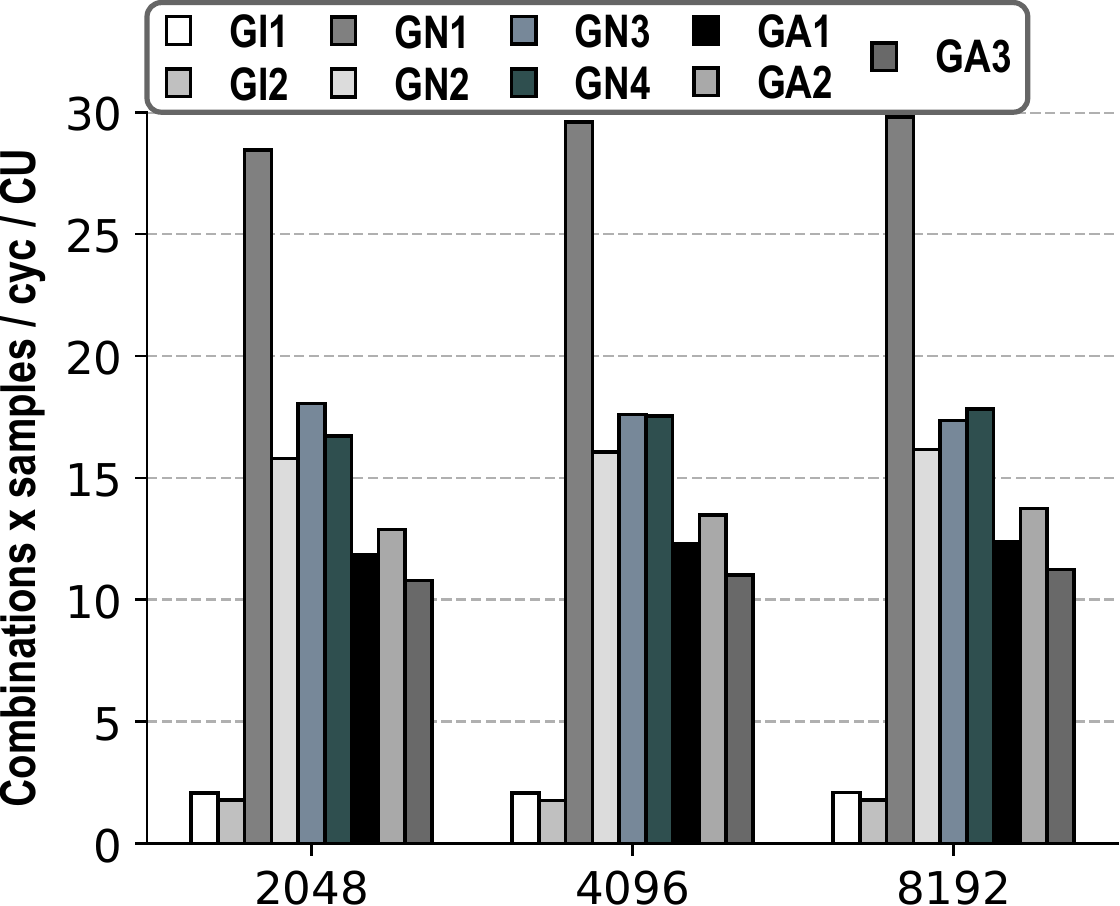}
	    \caption{Processed elements per cycle and \\ per compute unit.}
	    \label{fig:gpu_cyc_core}
    \end{subfigure}%
    \begin{subfigure}[t]{0.32\textwidth}
        \centering
	    \includegraphics[width=1\textwidth]{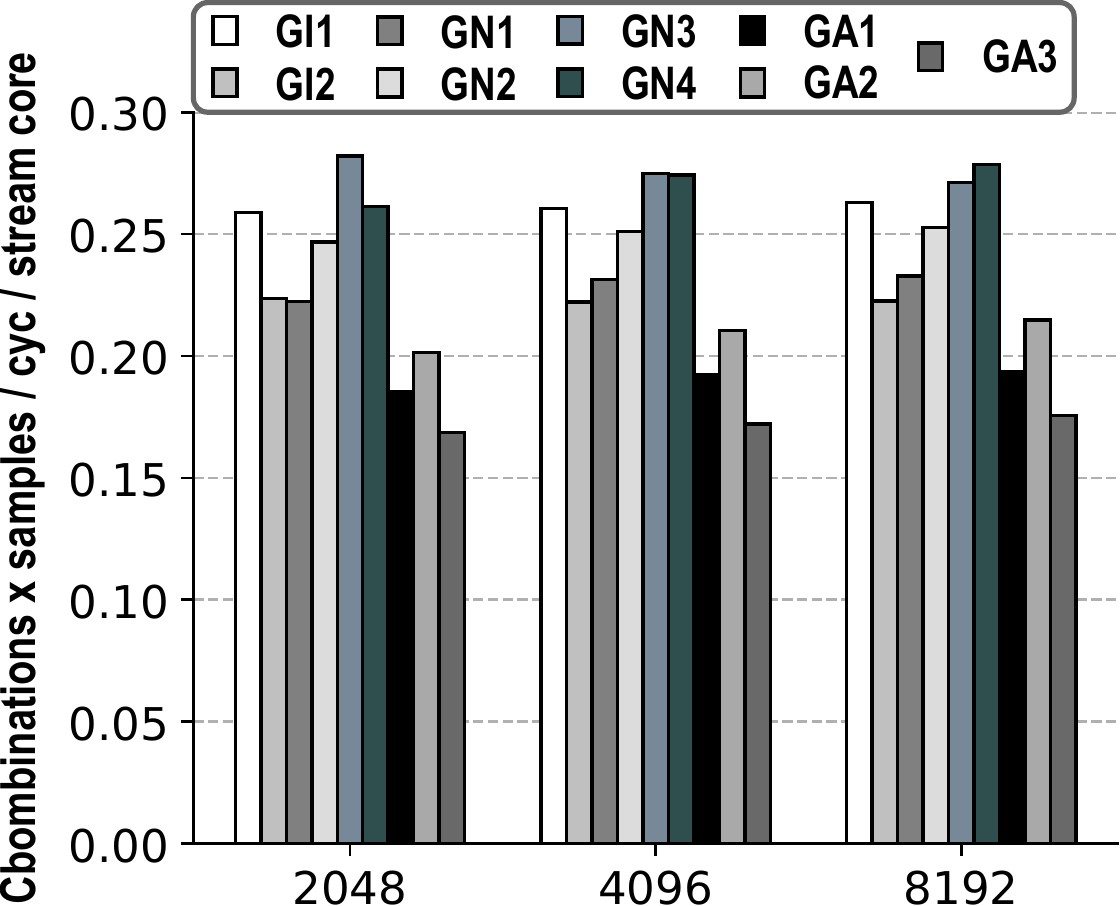}
	    \caption{Processed elements per cycle and per stream core.}
	    \label{fig:gpu_cyc_core_vec}
    \end{subfigure}%
    \caption{\ac{GPU} performance evaluation for different number of \acp{SNP} (2048, 4096, 8192) and 16384 samples.}
    \label{fig:gpu_results}
\end{figure*}

The performance as the number of processed elements per second and per core is presented in Figure~\ref{fig:cpu_sec_core}. As observed, the maximum performance is achieved by CI3 when using AVX512 vector intrinsics. For example, for 8192 \acp{SNP}, AVX512 CI3 attains a performance around 15.4 Giga combs. x samples / s / core, \textit{i.e.}, 2.5x and 4.8x higher than CI1 and AVX512 CI2, respectively. For the same data set and compared to the AMD \acp{CPU}, AVX512 CI3 delivers a performance per core 4x and 3x higher than CA1 and CA2, respectively. The higher performance on CI3 arises from the support of vector \textit{POPCNT} instructions introduced in the Ice Lake SP micro-architecture, which allows to fully use the vector capabilities of the processor when performing epistasis detection. For the remaining systems, the lack of vector \textit{POPCNT} instructions imposes the utilization of scalar \textit{POPCNT}, reducing performance. When using AVX512 instructions, CI2 (Skylake SP) achieves a performance lower than any other \ac{CPU}. As the Skylake SP micro-architecture requires the utilization of two extract instructions per each scalar \textit{POPCNT} when using AVX512 instructions, it imposes additional overheads. This effect is amplified by the frequency reduction that occurs in this architecture when using AVX512 instructions. In fact, when deploying the epistasis detection approach with AVX intrinsics on CI3 and CI2, the results show that the maximum performance is achieved by CI1 (6 Giga combs. x samples / s / core), followed by CA2 which attains a performance around 5 Giga combs. x samples / s / core. CI2, CI3 and CA1 achieve similar performance per core.  

When performing AVX instructions, the higher performance of CI1 and CA2 when compared to the other \acp{CPU} mainly arises from the higher frequency of these systems (see Table~\ref{tab:cpu_systems}). When considering the performance as the elements processed per cycle and per core (Figure~\ref{fig:cpu_cyc_core}), the approach vectorized with AVX intrinsics attains similar performance in all devices. In the case of CA1 and CA2, the increase in the vector width from 128-bit to 256-bit between Zen and Zen2 micro-architectures did not benefit the execution of three-way epistasis detection, due to the lack of vector \textit{POPCNT} instructions, resulting in similar performance. Similarly to the performance per second and per core, AVX512 execution on CI3 achieves the highest performance per cycle, \textit{i.e.}, approximately 3.8x higher than all the remaining \acp{CPU}.  

To assess the utilization efficiency of the vector units on the \acp{CPU}, the performance as the number of element processed per cycle and per core is also scaled to the number of elements of 32-bit that fit in the vector width of each \ac{CPU} (Figure~\ref{fig:cpu_cyc_core_vec}). The maximum performance is attained by CA1 and AVX512 CI3 and around 0.4. While in CA1 the high vector occupancy is due to its reduced vector width coupled with scalar \textit{POPCNT} instructions, for AVX512 CI3 this indicates that the usage of vector \textit{POPCNT} instructions allows to achieve a good occupancy of the vector units of Intel Ice Lake SP. The performance in CA2 is half of the obtained in CA1 demonstrating that the lack of vector \textit{POPCNT} instructions does not allow to fully use the wider vector width contained in the Zen2 micro-architecture, resulting in poor vector efficiency. For the same reason, CI1 achieves a performance up to 2.4x higher than CI2, since due to the lack of vector \textit{POPCNT}, it is not possible to take advantage of the increase in the vector width from 256-bit to 512-bit in Skylake SP. 

\subsection{GPU Evaluation}

The performance results on Intel, NVIDIA and AMD \acp{GPU} are presented in Figure~\ref{fig:gpu_results}. The configuration \textless$B_{Sched}, B_S$\textgreater~used in the best epistasis detection approach on the \ac{GPU} was defined empirically in order to maximize the performance in each device. The results presented in Figure~\ref{fig:gpu_results} were obtained with the following parameters: \textless256, 64\textgreater~for GI1 and GI2, \textless256, 32\textgreater~for GN1 and GA3, \textless256, 64\textgreater~for GN2, GN3 and GN4, and \textless128, 64\textgreater~for GA1 and GA2.  

\begin{table*}[]
\centering
\caption{Comparison with state-of-the-art approaches (* time estimated by assuming the same performance from the data set with 10000 \acp{SNP} and 1600 samples).}
\label{tab:soa_results}
\resizebox{0.99\linewidth}{!}{%
\begin{tabular}{@{}ccccccc@{}}
\toprule
SoA Work                           & SNPs                   & Samples               & Device                         & \begin{tabular}[c]{@{}c@{}}Performance of SoA Work\\ {[}Giga Combinations x samples / s{]}\end{tabular} & \begin{tabular}[c]{@{}c@{}}Performance of This Work\\ {[}Giga Combinations x samples / s{]}\end{tabular} & Speedup \\ \midrule

\multirow{8}{*}{MPI3SNP~\cite{doi:10.1177/1094342019852128}}       & \multirow{4}{*}{10000} & \multirow{4}{*}{1600} & NVIDIA Titan V                 & 663.4    & \textbf{1085.7}       &   1.64$\times$                                                                        \\
                               &                        &                       & NVIDIA Titan RTX               & 716.9   &      \textbf{1069.9}    &   1.49$\times$                                                                      \\
                               &                        &                       & (2x) Intel\textregistered~Xeon\textregistered~Platinum 8360Y & 38.8   &   \textbf{224.4}      &   5.78$\times$                                                                         \\
                               &                        &                       & AMD EPYC 7302P                 &      11.7      &      \textbf{67.1}    &   5.74$\times$                                                                    \\ \cmidrule(l){2-7} 
   & \multirow{4}{*}{40000} & \multirow{4}{*}{6400} & NVIDIA Titan V                 & 570.7    &   \textbf{1892.1}  &  3.31$\times$                                                                             \\
                               &                        &                       & NVIDIA Titan RTX               &   573.6  &    \textbf{2170.3}  &  3.78$\times$                                                                                 \\
                               &                        &                       & (2x) Intel\textregistered~Xeon\textregistered~Platinum 8360Y &   * ($\approx$ 20 days)  &        \textbf{818.3}  & $\approx$ 21.09$\times$                                                                               \\
                               &                        &                       & AMD EPYC\texttrademark~7302P                 &  * ($\approx$ 67 days)   & \textbf{* ($\approx$ 10 days)}       &      $\approx$ 6.70$\times$                        

\\ \midrule
\multirow{5}{*}{\cite{jsspp}} & \multirow{5}{*}{8000}  & \multirow{5}{*}{8000} 
& NVIDIA Titan Xp                & \textbf{1443.0}  &       1279.9        &  0.89$\times$                               \\
&                        &                       & NVIDIA Titan V                 & 1876.0  &    \textbf{1936.0}      & 1.03$\times$  \\
&                        &       & NVIDIA Titan RTX                &  2140 &       \textbf{2239}        &  1.05$\times$                                                           \\
&                        &       & NVIDIA A100 (250W)                &  2694 &      \textbf{2732}        &  1.01$\times$ \\
&                        &       & AMD Instinct Mi100                &  N/A &      \textbf{2249}        &  N/A
                                                             \\ \midrule
\multirow{2}{*}{\cite{10.1007/978-3-030-57675-2_38}}       & \multirow{2}{*}{1000}  & \multirow{2}{*}{4000} & Intel\textregistered~Graphics UHD P630        & 5.9    &    \textbf{62.3} &  10.56$\times$                                                                              \\
                               &                        &                       & Intel\textregistered~Core\texttrademark~i7-8700K            & 2.9  &   \textbf{30.3}  &   10.45$\times$                                                                                                                                     
                                                   
                                                  \\ \bottomrule
\end{tabular}%
}
\end{table*}

Regarding the performance as the elements processed per second and per compute unit (Figure~\ref{fig:gpu_sec_core}), the highest performance is achieved by GN1 (Titan Xp). For example, for 2048 \acp{SNP}, GN1 achieves a performance per compute unit 2x higher than GN2 (Titan V), 1.4x higher than GN3 (Titan RTX) and 1.9x higher than GN4 (A100). As shown in Table~\ref{tab:gpu_systems}, the higher performance per compute unit of GN1 arises from the higher \textit{POPCNT} throughput per compute unit on Titan Xp (32 \textit{POPCNT} per cycle) in comparison to Titan V, Titan RTX and A100 (16 per cycle)~\cite{nvidia_prog_guide}. For the AMD \acp{GPU}, GA1 (Mi100) and GA2 (Radeon Pro VII) attain a similar performance and lower than GA3 (RX 6900 XT). Although the throughput of \textit{POPCNT} per compute unit on GA1 and GA2 (12 per cycle) is higher than the one on GA3 (10 per cycle), the higher frequency of GA3 allows this \ac{GPU} to attain higher performance than GA1 and GA2. Similarly for Intel \acp{GPU}, the higher frequency of GI2 results in slightly higher performance than GI1, as both \acp{GPU} have the same \textit{POPCNT} throughput per compute unit.

The effect of the frequency can be isolated by considering the performance as the elements per cycle and per compute unit (Figure~\ref{fig:gpu_cyc_core}). While the maximum performance is still achieved by GN1, the difference between GN2, GN3 and GN4 reduces, indicating that the main differentiating factor between Titan V, Titan RTX and A100 when performing epistasis is the higher frequency of Titan RTX. For AMD, the performance of GA1 and GA2 is higher than GA3, corroborating with the \textit{POPCNT} information in Table~\ref{tab:gpu_systems}. For Intel \acp{GPU}, without the effect of frequency on performance, it is possible to verify that the performance per compute unit is similar on both devices, with GI1 having a slight advantage over GI2. 

To assess the occupancy of the computing units, the performance is also scaled to the number of stream cores (Figure~\ref{fig:gpu_cyc_core_vec}). Intel and NVIDIA \acp{GPU} achieve similar performance (between 0.27 and 0.23 combs. x samples / cyc / stream core), indicating that the ratio of available units for \textit{POPCNT} and the total number of stream cores is similar on these \acp{GPU}. As the architecture of the stream cores of GN3 and GN4 are similar, they achieve similar maximum performance and around 0.27 combs. x samples / cyc / stream core.  The lower  number of stream cores that support \textit{POPCNT} on AMD \acp{GPU} results in lower occupancy than Intel and NVIDIA achieving a performance between 0.175 (GA3) and 0.21 (GA1) combs. x samples / cyc / stream core.

\subsection{Comparison between CPUs and GPUs}

By comparing the performance of the \acp{CPU} scaled to the number of cores and vector width (Figure~\ref{fig:cpu_cyc_core_vec}) and the \ac{GPU} performance per cycle and per stream core (Figure~\ref{fig:gpu_cyc_core_vec}), it is possible to observe that the performance of CI1, CA1, CA2 and AVX512 CI3 is similar to the performance of \acp{GPU}. Hence, the higher performance on \acp{GPU} relatively to the \acp{CPU} when evaluating three-way gene interactions mainly arises from the high number of stream cores. The sole exception is CA1, which attains high performance due to the reduced vector width coupled with the scalar \textit{POPCNT}. This shows that for a \ac{CPU} to achieve an overall performance close to the \acp{GPU} when performing three-way epistasis detection, it must feature wider vector units and with a higher core count, increasing the number of elements that can be processed simultaneously.

Since the best approaches on \ac{CPU} and \ac{GPU} are compute bound, their performance is highly dependent on the performance of \textit{POPCNT}. For this reason, to maximize \ac{CPU} performance, it is crucial to support vectorized \textit{POPCNT}, in order to take full advantage of the vector units. For the \ac{GPU}, the same insights can be derived for the number of stream cores that support \textit{POPCNT} instructions, since as more cores support it, the higher the performance is expected to be in these devices. The clock frequency and number of cores are also highly relevant when evaluating of three-way gene interactions, as shown in Figures~\ref{fig:cpu_results} and~\ref{fig:gpu_results}. Due to its compute bound nature, the performance of the application is expected to scale linearly with the number of cores/compute units and frequency. Thus, devices with a higher core count and more compute units coupled with high frequency are better suited for high-performance execution on current \acp{CPU} and \acp{GPU}. 

Another solution to speed up three-way epistasis detection is to use heterogeneous systems with \ac{CPU}+\ac{GPU}. However, most of the evaluated \acp{CPU} attain a performance much lower than the one achieved with \acp{GPU}, which translates to a poor performance improvement for heterogeneous approaches. For example, while the NVIDIA Titan RTX (GN3) achieves a maximum overall performance around 2200 Giga combs. x samples / s, the Intel 8700K (CI1) and AMD EPYC 7601 (CA1) only achieve a maximum performance around 36.5 and 241 Giga combs. x samples / s, respectively. From the \acp{CPU} considered in this work, the Intel Ice Lake SP (CI3) is the most suitable \ac{CPU} to be incorporated in a heterogeneous approach, since it is able to deliver a performance around 1100 Giga combs. x samples / s, \textit{i.e.}, half of the NVIDIA Titan RTX. A heterogeneous solution with CI3+GN1 would be expected to achieve a performance up to 3300 Giga combs. x samples / s.

The experimental results for the \acp{GPU} also show that the portability provided by the proposed approaches allows to achieve high-performance execution on devices with different characteristics. For example, when comparing AMD Mi100 and NVIDIA Titan RTX \acp{GPU}, AMD Mi100 is able to deliver higher performance (around 2.5 Tera combs. $\times$ samples / s) than the one offered by NVIDIA Titan RTX (around 2.3 Tera combs. $\times$ samples / s). Of the tested GPUs, only the most recent NVIDIA \ac{GPU}, \textit{i.e.}, NVIDIA A100 is able to surpass the performance offered by AMD Mi100, achieving around 2.7 Tera combs. $\times$ samples / s. In this scenario, the epistasis detection approaches proposed in this work have the potential to efficiently support multiple different architectures, as well as current and future devices.

On the other hand, while NVIDIA and AMD \acp{GPU} are the most suitable devices to attain high performance, since they are able to deliver an overall performance above 2000 Giga combs. x samples / s, from an efficiency aspect, the best device is the Intel Iris Xe Max (GI2). Although this \ac{GPU} only delivers an overall performance up to 282.1 Giga combs. x samples / s, this is achieved with a TDP of 25 W, attaining  an estimated efficiency of 11.3 Giga combs. x samples / J. In comparison, NVIDIA Titan RTX delivers a performance of 2200 Giga combs. x samples / s with a TDP of 280W, \textit{i.e.}, an efficiency of 7.9 Giga combs. x samples / J. In this scenario, the Intel Iris Xe MAX \ac{GPU} is the most appropriate device to efficiently verify if a patient has a high risk of developing a certain disease on personalized healthcare services, by knowing \textit{a priori} which \acp{SNP} to evaluate. For exploratory analysis on entire data sets, high-performance devices are the best choice.

\subsection{Comparison with state-of-the-art}

The performance of the best approach for three-way epistasis detection proposed in this work is also compared against state-of-the-art works on three-way epistasis, namely MPI3SNP~\cite{doi:10.1177/1094342019852128}, and the works presented in~\cite{jsspp} and~\cite{10.1007/978-3-030-57675-2_38}. The MPI3SNP results were obtained by executing the application on the experimental platforms used in this work, with the data sets contained in the project repository\footnote{https://github.com/UDC-GAC/mpi3snp/wiki/Sample-files}. Similarly, the results for~\cite{jsspp} were obtained through experimental evaluation on the considered devices. The values from~\cite{10.1007/978-3-030-57675-2_38} are directly obtained from the respective manuscript.

As observed in Table~\ref{tab:soa_results}, the best approach for three-way gene interactions proposed in this work attains higher performance than MPI3SNP. For the data set containing 10000 \acp{SNP} and 1600 samples, the proposed approach attains a performance 1.6x and 1.5x higher in Titan V and Titan RTX, respectively. For the Intel 8360Y and AMD 7302P, the performance gains are up to 5.8x. As for the data set with 40000 \acp{SNP} and 6400 samples, the approach proposed in this work attains a performance 3.3x higher than MPI3SNP for Titan V, and 3.8x higher for Titan RTX. For the \acp{CPU}, due to the unreasonable execution time of MPI3SNP in Intel 8360Y and AMD 7302P, we assumed that for this data set, the performance of MPI3SNP is equal to the one obtained for data set with 10000 \acp{SNP} and 1600 samples. In this scenario, it is expected that the best approach considered in this work achieves performance gains up to 21.1x in the Intel 8360Y. In fact, while our approach was able to process the data set containing 40000 \acp{SNP} and 6400 samples in approximately 1 day on an Intel 8360Y, the MPI3SNP is expected to take approximately 20 days on the same machine.

While maintaining a focus on performance portability, our approach is still able to be on par with a highly optimized and hand-tuned CUDA algorithm for three-way epistasis detection, while being deployed using DPC++. This is observed particularly in the most recent NVIDIA architectures, where our approach provides slight performance improvements over the work in~\cite{jsspp} for Titan V, Titan RTX and A100. Furthermore, when considering the performance of our approach on the AMD Instinct Mi100, results show that supporting portability between different architectures can lead to higher performance. While the work in~\cite{jsspp} cannot be run on this device, the performance obtained by our approach is 1.56$\times$, 1.20$\times$ and 1.05$\times$ higher when compared to~\cite{jsspp} in NVIDIA Titan Xp, Titan V and Titan RTX, respectively. These results constitute, to the best of our knowledge, the highest performance obtained for a three-way exhaustive epistasis detection approach on AMD GPUs. Only the most recent NVIDIA GPU (A100) is able to surpass the performance of the AMD Mi100 by 1.2$\times$. For the Intel Gen9.5 \ac{GPU}, the proposed approach attains a performance around 62.3 Giga combs. x samples / s, \textit{i.e.}, 10.6x higher than the performance obtained in~\cite{10.1007/978-3-030-57675-2_38}, while for Intel 8700K, the performance of the proposed method is 10.4x higher than the one obtained in~\cite{10.1007/978-3-030-57675-2_38}.

\section{Conclusions and Future Work}
\label{sec:conc}

To achieve the goals of personalized healthcare, the development of bioinformatics applications that relate patient genetic data with the risk of disease development is essential. Exhaustive epistasis detection constitutes one such application, relying on large biological data sets and complex operations to identify gene interactions. This work proposes a set of approaches for three-way exhaustive epistasis detection on modern \acp{CPU} and \acp{GPU}, that employ several optimizations to suit a range of target architectures. Using insights from \ac{CARM}, the most adequate approaches were identified, and an exhaustive study on 5 \acp{CPU} and 8 \acp{GPU} from all main manufacturers was presented. This allowed to identify the main features relevant to obtain high-performance in bioinformatics applications. Moreover, the proposed approaches were able to obtain higher performance than state-of-the-art works in all platforms, achieving speedups of up to 10.6$\times$.

Future directions include the evaluation of several micro-architectures from the perspective of optimization goals, such as, power consumption and energy efficiency, as well as the inclusion of DVFS techniques to further improve the efficiency of bioinformatics applications.

\bibliographystyle{IEEEtran}
\bibliography{IEEEabrv,bibliography}

\end{document}